\documentclass[twocolumn,american,aps,pra,showkeys,showpacs,floatfix,reprint]{revtex4-1}
\usepackage{lmodern}
\usepackage[T1]{fontenc}
\usepackage[koi8-r,latin9]{inputenc}
\usepackage{geometry}
\usepackage{units}
\usepackage{amsmath}
\usepackage{amssymb}
\usepackage{graphicx}
\usepackage{setspace}

\makeatletter

 \@ifundefined{textcolor}{}
 {%
   \definecolor{BLACK}{gray}{0}
   \definecolor{WHITE}{gray}{1}
   \definecolor{RED}{rgb}{1,0,0}
   \definecolor{GREEN}{rgb}{0,1,0}
   \definecolor{BLUE}{rgb}{0,0,1}
   \definecolor{CYAN}{cmyk}{1,0,0,0}
   \definecolor{MAGENTA}{cmyk}{0,1,0,0}
   \definecolor{YELLOW}{cmyk}{0,0,1,0}
 }

\makeatother

\usepackage{babel}
\begin{document}

\title{On some links between quantum physics and gravitation}

\author{Aleksey V. Ilyin}

\email{a.v.ilyin@mail.mipt.ru}

\selectlanguage{american}%

\affiliation{Moscow Institute of Physics and Technology}

\date{\today}
\begin{abstract}
It is widely believed that quantum gravity effects are negligible
in a conventional laboratory experiment because quantum gravity should
play its role only at a distance of about Planck's length ($\sim10^{-33}$
cm). Sometimes that is not the case as shown in this article. We discuss
two new ideas about quantum physics connections with gravity. First,
the Hong-Ou-Mandel effect relation to quantum gravity is examined.
Second, it is shown that the very existence of gravitons is a consequence of quantum statistics. Moreover, since the Bose-Einstein
statistics is a special case of Compound Poisson Distribution, it
predicts the existence of an infinite family of high-spin massless
particles. 
\end{abstract}



\maketitle

\section{INTRODUCTION\label{sec:Intro}}

It is a common knowledge that gravity plays no role in conventional
quantum phenomena because the gravitational interaction is negligible
if compared to electromagnetic or strong interaction. The role of
gravity in quantum mechanics is usually illustrated by the following
example~\cite{Gorelik}: radiative decay of an excited state of an
atom due to electromagnetic interaction takes about $10^{-10}$ sec,
whereas the same decay due to gravitational interaction must go on
for about $10^{30}$ years, which is much longer than the age of the
Universe.

Such an insignificant role of gravity in atomic processes, however,
does not prevent it from determining the motion of celestial bodies
and playing a prominent role in people's daily life. Gravity acts
in the first place where all other forces are compensated or absent.
But similar situation is also possible, as shown below, in some quantum
processes. This requires that no other interactions, such as electromagnetic,
interfere with quantum gravity effects.

In Section~\ref{sec:HOM-Effect}, the Hong-Ou-Mandel experiment~\cite{HOM}
is analyzed for quantum gravity effects. We start from a short description
of the HOM-effect, then we prove that coalescent photons observed
in the experiment do not obey the Maxwell's equations for electromagnetic
field, and finally we discuss what equations may be used to describe
coalescent photons. Strong reasons are presented in support of conclusion
that coalescent photons obey the Einstein's equations of general relativity.

Another example of quantum physics connection to gravity is considered
in Section~\ref{sec:BE_&_Gravity}, which shows that quantum statistics
necessarily predicts the existence of gravitons and an infinite family
of high\nobreakdash-spin massless particles provided there is a spin-one
massless particle (photon). This conclusion follows from the fact
that the Bose-Einstein (BE) statistics coincides with negative binomial
distribution, which, as shown in \cite{Compound_Distribution}, is
a special case of Compound Poisson Distribution. Wherefrom it follows
that the BE statistics describes a stream of composite random events,
each composite event consisting of random number of elementary events.
In quantum optics, an elementary event may only be a registration
of single photon, so that a composite event is a registration of photon
cluster, i.\,e. a high-spin massless particle.

Main results of this work are discussed in Section~\ref{sec:Conclusions}.

\section{HOM-EFFECT\label{sec:HOM-Effect}}

\begin{figure}[ht]
\centering{}\includegraphics[width=5cm]{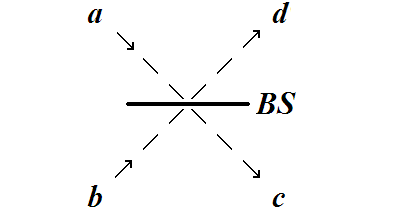}\caption{{\small Hong-Ou-Mandel effect: two identical photons $a$ and~$b$,
upon impinging on a beam splitter (BS), are found either in the output
mode $c$ or in the output mode~$d$. Therefore, such photons are
deemed coalescent.\label{fig:HOM}}}
\end{figure}

In the famous experiment by Hong, Ou and Mandel \cite{HOM} two identical
photons $a$ and $b$ impinge on two different input ports of a symmetric
beam splitter (Figure~\ref{fig:HOM}). After interacting with the
beam splitter both photons are found in a single output mode - either
in mode $c$ or in mode $d$. Such photon pairs in a single mode were
called the coalescent photons~\cite{Kaige Wang}.

Two coalescent photons show a remarkable behavior: if they are allowed
to fall on a second beam splitter (not shown) then they are either
together transmitted or together reflected at the beam splitter~\cite{Stability_of_clusters}.
In other words, coalescent photons behave as if they were a single
quantum object that is never separated into two different photons
when interacting with a semi-transparent mirror. Therefore, it is
convenient to call such an object a two-photon cluster~\cite{Ilyin-CP_in_BR}.

Apart from two coalescent photons, three and even more coalescent
photons were observed in different experiments~\cite{Two-photons,Three-photons,Four-photons,Six-photons}.
For that reason, $N$ coalescent photons will be termed here as the
$N$\nobreakdash-photon cluster.

\subsection{Photon clusters do not obey Maxwell's equations}

It is important that $N$\nobreakdash-photon cluster field does not obey the Maxwell's equations for electromagnetic field. Indeed,
the momentum of $N$-photon cluster is $N$ times the momentum of
a single photon. According to de~Broglie's formula $\lambda=\tfrac{h}{p}$,
the wavelength of a particle is determined by its momentum. Therefore,
the wavelength of $N$\nobreakdash-photon cluster must be $N$ times
smaller than the wavelength $\lambda$ of one photon: 
\begin{equation}
\lambda_{N}=\dfrac{\lambda}{N},\label{eq:Lambda_N}
\end{equation}
 where $\lambda_{N}$ is the wavelength of $N$-photon cluster. Reduction
in the wavelength \eqref{eq:Lambda_N} when photons stick together
was repeatedly confirmed in multiphoton experiments involving up to
six coalescent photons~\cite{Two-photons,Three-photons,Four-photons,Six-photons}.
Therefore, \eqref{eq:Lambda_N} may be considered a reliably established
rule.

Phase velocity of a particle is determined by its frequency and wavelength
$v_{ph}=\nu\lambda$. Photons do not interact, so when they stick
together their energy $\varepsilon=h\nu$ does not change, hence,
their frequency remains constant. So, when photons stick together,
their phase velocity changes as the wavelength. Phase velocity of
a single photon equals the velocity of light $v_{ph}=\nu\lambda=c$.
Therefore, phase velocity of $N$-photon cluster, due to \eqref{eq:Lambda_N},
is 
\begin{equation}
v_{ph}=\nu\lambda_{N}=\dfrac{c}{N},\label{eq:Phase_velocity}
\end{equation}
 which coincides with the speed of light only for single photons.
Thus, phase velocity of $N$-photon cluster field in vacuum is $N$
times less than the speed of light.

The Maxwell's equations for a plane wave in vacuum, however, cannot
produce any phase velocity other than the speed of light. Therefore,
the $N$-photon cluster field does not obey the Maxwell's equations.

In this regard, a natural question arises, what equations may describe
a photon cluster field?

\subsection{Search for suitable equations}

Since a photon has spin $s=1$, a two-photon cluster that consists
of two indistinguishable photons has spin~$s=2$. It is a common
belief that a spin-2 particle must be described by a second rank tensor
field in contrast to a spin-1 particle that is described by a vector
field. For this reason, a two-photon cluster must be described by
a second rank tensor field. Maxwell's equations, of course, are only
suitable for tensor fields of rank one. That is another reason why
photon clusters do not obey Maxwell's equations.

It should be noted that L.~Landau was the first to prove (in 1948)
that two photons in a mode should be described by a second rank tensor
field~\cite{Landau}. Such tensor field was then believed to obey
some linear differential equations other than the Maxwell's equations.

A search for equations that could describe spin\nobreakdash-2 massless
fields was conducted within the field-theoretic approach during almost
the entire second half of the 20th century.

At the beginning, different linear equations were studied. Many options
were considered but later a theorem was proved that any system of
linear equations for a second rank tensor field would inevitably produce
insurmountable contradictions if considered on the curved background,
i.\,e. in the presence of external gravitational field~\cite{Linear Spin-2 fields}.
Thus, no linear equations for spin\nobreakdash-2 fields are possible.

Then various nonlinear equations for the tensor field were investigated.
Several different systems of nonlinear equations were found that met
all the requirements and did not lead to any contradictions. However,
later it was proved in \cite{Nonlinear Spin-2 fields} and \cite{Nonlinear Spin-2 Deser}
that any such system of nonlinear equations could be reduced to the
Einstein's equations of general relativity by appropriate coordinate
transformations. Thus, all these nonlinear equations were actually
the Einstein's equations in various, sometimes exotic, coordinate
systems. A detailed discussion of this issue can be found in~\cite{Sokolowski}.

The main result of these studies may be given as follows: a consistent
description of massless rank\nobreakdash-2 tensor field is only possible
within the Einstein's equations. Consequently, a two-photon cluster
field must obey the Einstein's equations, as it is a spin\nobreakdash-2
massless field.

\subsection{Another approach}

The same result can be obtained using a different method. There is
a simple quantum principle, which states that if two quantum particles
are \emph{identical} then they must have the same quantum numbers.
For example, two identical particles must have the same mass, charge,
spin, etc. The reverse is also true: if two particles have the same
quantum numbers then they are identical.

A two-photon cluster is a spin\nobreakdash-2 massless particle. The
same quantum numbers are known to characterize gravitons. Consequently,
a two-photon cluster is identical to graviton and must, therefore,
obey the same Einstein's equations that are believed to describe gravitons.
This conclusion can be formulated in short: there is only one spin\nobreakdash-2
massless particle in nature~\cite{Nonlinear Spin-2 Deser}.

That means that in the 1987 Hong\nobreakdash-Ou\nobreakdash-Mandel
experiment \cite{HOM}, \emph{optical frequency gravitons} were registered
at the output ports of the beam splitter. Such gravitons very much
differ from those present in the Earth's gravitational field.

Indeed, experimentally observed gravitons are \emph{real} particles
while a stationary gravitational field contains \emph{virtual} gravitons
like electrostatic fields contain virtual photons. The difference
between virtual and real particles is tremendous -- a virtual particle
cannot propagate to infinity while real particles may carry energy
away to an infinite distance. In addition, virtual gravitons in the
Earth's field have wavelengths that are determined by a typical distance
over which the Earth's gravitational field extends, that is about
tens of thousands of kilometers. Gravitons in the HOM-type experiments
have a wavelength less than a micron, i.\,e. at least 12 orders of
magnitude smaller than wavelengths of virtual gravitons in the Earth's
field. Finally, gravitons in the Earth's field are unpolarized particles,
while in the experiments polarized gravitons are observed. These differences
explain why gravitons observed in HOM-type experiments are so different
from the virtual gravitons that make up the Earth's gravitational
field.

Any deviation from an ideal alignment in a HOM-type experiment will
result in a mixture of tensor and vector fields propagating in output
modes $c$ and $d$ (Figure~\ref{fig:HOM}). In this case the efficiency
of photon coalescence will be less than 100\% and single photons will
sometimes appear in two output modes. Probability of such event is
defined by the vector field amplitude. That is the usual situation
in such experiments.

If a two-photon cluster is identical to graviton then, within this
paradigm, a 3-photon cluster should be a kind of coalescent state
of photon and graviton, while a 4-photon cluster is the coalescent
state of two gravitons. Since multiphoton clusters of various ranks
have been observed in several experiments~\cite{Two-photons,Three-photons,Four-photons,Six-photons},
it seems probable that there is a great variety of high-spin massless
fields, representing different combinations of coalescent photons
and gravitons. In the next Section, this conjecture will be supported
by a sound validation within the framework of quantum statistics.

\section{QUANTUM STATISTICS AND GRAVITY\label{sec:BE_&_Gravity}}

The BE statistics is usually considered in a single cell of phase
space (that corresponds to one coherence volume): 
\begin{equation}
p_{n}(1)=\dfrac{w^{n}}{\left(1+w\right)^{n+1}},\label{eq:BE_in_single_bin}
\end{equation}
 where $p_{n}(1)$ is the probability that $n$ photons are in one
coherence volume, $w$ is the average number of photons per coherence
volume.

For arbitrary phase-space volume $\tau$, the BE statistics has the
form: 
\begin{equation}
p_{n}(\tau)=C_{\tau+n-1}^{n}\frac{w^{n}}{\left(1+w\right)^{n+\tau}},\label{eq:BE_in_tau_bins}
\end{equation}
 where $C_{\tau+n-1}^{n}$ is a binomial coefficient:

\begin{eqnarray}
C_{\tau+n-1}^{n}=\dfrac{\left(\tau+n-1\right)\,!}{n\,!\left(\tau-1\right)\,!} & = & \frac{\tau(\tau+1)\ldots(\tau+n-1)}{n\,!}.\nonumber \\
\label{eq:Comb}
\end{eqnarray}

Formula \eqref{eq:BE_in_tau_bins} was derived by Leonard Mandel for
an integer number of cells~\cite{BE_by_Mandel}. It was later shown
in \cite{GBD} that the Mandel's formula \eqref{eq:BE_in_tau_bins}
is valid for an arbitrary volume~$\tau$ including nonintegral number
of coherence volumes. It is clear that the last expression in \eqref{eq:Comb}
makes sense for any positive value of~$\tau>0$. If $\tau=1$ then
\eqref{eq:BE_in_tau_bins} becomes the usual expression \eqref{eq:BE_in_single_bin}
for the BE statistics in a single cell.

Coalescence of particles in the BE statistics is a formal consequence
of the fact that the BE statistics \eqref{eq:BE_in_tau_bins} coincides
with a negative binomial distribution, which has the form 
\begin{equation}
p_{n}(\tau)=C_{\tau+n-1}^{n}p^{\tau}\left(1-p\right)^{n},\label{eq:NBD}
\end{equation}
 where coefficients $C_{\tau+n-1}^{n}$ are defined in \eqref{eq:Comb},
parameters $p$ and $\tau$ must satisfy $0\leq p\leq1$ and $\tau>0$,
respectively.

If parameter 
\begin{equation}
p=\dfrac{1}{1+w}\label{eq:p(w)}
\end{equation}
 then \eqref{eq:NBD} coincides with the BE statistics~\eqref{eq:BE_in_tau_bins}.

According to \cite{Compound_Distribution}, probability distribution
\eqref{eq:NBD} is a special case of \emph{Compound Poisson Distribution},
so that \eqref{eq:NBD} describes a flow of random composite events.
It was established in \cite{Compound_Distribution} that these composite
events obey Poisson statistics 
\begin{equation}
g_{k}(\tau)=\dfrac{(\eta\tau)^{k}}{k\,!}e^{-\eta\tau},\label{eq:Cluster_Statistics}
\end{equation}
 where $\eta$ is the average number of composite events per unit
volume 
\begin{equation}
\eta=\ln\dfrac{1}{p},\label{eq:Feller_eta}
\end{equation}
 while probability $f_{k}$ that a composite event consists of $k$
elementary events is given by a logarithmic distribution: 
\begin{equation}
f_{k}=\dfrac{(1-p)^{k}}{k\eta}.\label{eq:Feller_fk}
\end{equation}

In the case of photon statistics, an elementary event may only be
the registration of one photon by an ideal detector. Then a composite
event will be a simultaneous registration of several photons, i.\,e.
of photon cluster.

The above results, obtained in \cite{Compound_Distribution} from
the general theory of Compound Poisson Distribution with respect to~\eqref{eq:NBD},
were later derived in \cite{Ilyin-CP_in_BR} from the BE statistics
without resorting to Compound Poisson Distribution. Such a new derivation
provides an independent confirmation of these results.

Due to \eqref{eq:p(w)}, equations \eqref{eq:Feller_eta} and \eqref{eq:Feller_fk}
take the form 
\begin{equation}
\eta=\ln(1+w),\label{eq:BE_eta}
\end{equation}
\begin{equation}
f_{k}=\dfrac{w^{k}}{k(1+w)^{k}\ln(1+w)}.\label{eq:BE_fk}
\end{equation}

In the case of quantum statistics, \eqref{eq:BE_eta} gives the average
number of photon clusters per coherence volume (if no distinction
is made between clusters of different ranks), while \eqref{eq:BE_fk}
is the probability that a photon cluster consists of $k$ photons,
$k=1,\,2,\,3,\,...$~.

In the BE statistics, the average number of photons per coherence
volume is 
\begin{equation}
w=\dfrac{1}{\exp\left(\beta\varepsilon\right)-1},\label{eq:Mean_ph_number}
\end{equation}
 where $\beta=\nicefrac{1}{kT}$, and $\varepsilon=h\nu$. Therefore,
distribution of photon clusters by rank \eqref{eq:BE_fk}, in view
of~\eqref{eq:Mean_ph_number}, is a function of radiation frequency
and blackbody temperature.

It follows from the above that quantum statistics predicts both the
existence of photon clusters and distribution of clusters by rank
starting from single photons, when $k=1$, to infinite cluster ranks
for $k\rightarrow\infty$. It follows from~\eqref{eq:BE_fk}, however,
that the relative frequency of occurrence of high-rank clusters rapidly
falls with increasing the rank. In other words, single photons are
most frequently met in the blackbody radiation while two-photon clusters
are less frequent,~etc. This issue was investigated in greater detail
in \cite{Ilyin-CP_in_BR} where spectra of cluster radiation in blackbody
cavity were also found.

Thus, the mathematical fact that the BE statistics is actually a Compound
Poisson Distribution implies that there are photon clusters of
various ranks in blackbody radiation. 

If a two-photon cluster is really identical to graviton, as it follows
from Section~\ref{sec:HOM-Effect}, then the BE statistics predicts that the blackbody radiation contains gravitons and other high-spin massless fields corresponding to various $N$-photon clusters.

\section{CONCLUSIONS\label{sec:Conclusions}}

In this work it is shown that the Hong-Ou-Mandel effect is directly
related to quantum gravity. First, it is proven that coalescent photons
appearing in the HOM-effect do not obey the Maxwell's equations for
electromagnetic field. Second, strong arguments are given in favor
of conclusion that two coalescent photons (referred to as a two\nobreakdash-photon
cluster) must obey the Einstein's equations of general relativity.
Therefore, a two\nobreakdash-photon cluster is actually an optical
frequency graviton. For this reason, the Hong-Ou-Mandel experiment
\cite{HOM} and many other HOM-type experiments in the field of quantum
information are actually directly related to quantum gravity.

The idea behind this conclusion is simple: two identical photons,
due to their bosonic nature, may occupy the same quantum state that,
according to~\cite{Landau}, must be described by a second-rank tensor
field. Such tensor field, according to \cite{Nonlinear Spin-2 Deser,Nonlinear Spin-2 fields,Sokolowski},
can only obey the Einstein's equations of general relativity. 

In the second part of this work it is shown that the existence of
gravitons, as well as an infinite family of massless
fields with higher spins, is an inevitable consequence of quantum
statistics. The infinite hierarchy of high-spin massless fields appears
as a set of composite events in the Compound Poisson Distribution
that describes the Bose-Einstein statistics.

\end{document}